\documentstyle[sprocl]{article}

\def\Journal#1#2#3#4{{#1} {\bf #2}, #3 (#4)}

\def\CPC{\em Comp. Phys. Commun.}
\def\CPL{\em Chem. Phys. Lett.}
\def\EPL{\em Europhys. Lett.}
\def\JChP{\em J. Chem. Phys.}
\def\JCoP{\em J. Comp. Phys.}
\def\JPA{{\em J. Phys.} A}
\def\JPP{\em J. de Physique I (Paris)}
\def\JPSJ{\em J. Phys. Soc. Japan}
\def\JSP{\em J. Stat. Phys.}
\def\NAT{\em Nature}

\def\NPB{{\em Nucl. Phys.} B}
\def\PLA{{\em Phys. Lett.}  A}
\def\PLB{{\em Phys. Lett.}  B}
\def\PHA{{\em Physica} A}
\def\PRA{{\em Phys. Rev.} A}
\def\PRB{{\em Phys. Rev.} B}
\def\PRD{{\em Phys. Rev.} D}
\def\PRE{{\em Phys. Rev.} E}
\def\PRL{\em Phys. Rev. Lett.}
\def\ZPB{{\em Z. Phys.} B}

\def\be{\begin{equation}}
\def\ee{\end{equation}}
\def\bea{\begin{eqnarray}}
\def\eea{\end{eqnarray}}

\begin{document}

\title{BRIEF REVIEW OF MULTICANONICAL SIMULATIONS~\footnote{Published 
in the Proceedings of the International Conference on Multiscale 
Phenomena and Their Simulations (Bielefeld, October 1996), edited by 
F. Karsch, B. Monien and H. Satz (World Scientific, 1997).}}

\author{ BERND A. BERG~\footnote{E-mail: berg@hep.fsu.edu;
WWW: http://www.hep.fsu.edu/\~\,berg.}}

\address{Department of Physics, Florida State University,
\\Tallahassee, FL 32306, USA\\ and\\
Supercomputer Computations Research Institute,\\
Florida State University, Tallahassee, FL 32306, USA}


\maketitle\abstracts{Recent progress of simulations with non-canonical
weights is summarized.}
 
\section{Introduction}

One of the questions which ought to be addressed before performing a
large scale computer simulation is ``What are suitable weight factors 
for the problem at hand?'' It has been expert wisdom for quite a while
that Monte Carlo (MC) simulations with a-priori unknown weight factors
are feasible and deserve to be considered
\cite{ToVa77}. With focus on narrow classes of applications, this
idea was occasionally re-discovered, for instance \cite{Ba87}. But
it needed the work of Ref.\cite{BeNe91,BeNe92} to become a more widely
accepted idea in Statistical Physics as well as in Lattice Gauge
Theory.

The next section introduces systems with supercritical slowing down.
Section~3 reviews the multicanonical (MUCA) approach and briefly sketches 
related methods. Applications to first-order phase transitions are summarized 
in section~4 and those to systems with conflicting constraints in section~5,
before some conclusions are drawn in section~6. Some interesting topics,
like the dynamical-parameter method in $U(1)$ gauge theory~\cite{KeReWe95},
had to be omitted because of space limitations. For a MUCA calculation
of constraint effective potentials see Neuhaus in these proceedings.

\section{Supercritical Slowing Down}

The terminology ``supercritical slowing down'' is used to characterize 
canonical MC simulations which slow down exponentially fast with 
increasing system size. It is useful to distinguish {\it static} and
{\it dynamic} reasons for the slowing down. In the former case desired 
configurations have an exponentially small weight, whereas in the 
latter case their weight may still be large, but the dynamic process of 
reaching them deteriorates.
\medskip

{\it Static Examples:}

\begin{itemize}
\item Magnetic field driven first-order phase transitions:
Configurations with zero (or small) magnetic fields are
exponentially suppressed at low temperatures and they
exhibit domain walls.
\item Temperature driven first-order phase transitions:
Configurations with domain walls are exponentially suppressed.
\end{itemize}

{\it Dynamic Examples:}

\begin{itemize}
\item Low temperature transitions between magnetic states
(for instance the up-down states of the
Ising model below the Curie temperature, ...).
\item Transition between low temperature states in systems
with conflicting constraints: Spin glasses, the traveling
salesman problem, ...
\end{itemize}

\section{Multicanonical and Related Methods}

These methods try to overcome supercritical slowing down by
sampling with {\it unconventional weights}.

Canonical MC simulations perform importance sampling with respect 
to the Boltzmann weight
$$ w_B = e^{-\beta E} $$
Using reweighting, expectation values in a vicinity $(\to 0 ~{\rm for}~
V\to\infty)$ of the temperature $T=1/\beta$ are obtained.

{\it Multicanonical} refers to simulations which obtain expectation
values for a {\it temperature range}, which stays finite in the
limit $V\to\infty$.

Similarly, {\it multimagnetical} \cite{BeHaNe93} refers to simulations which 
give results for a certain range of the magnetic field, etc. ...~. The
cluster variant \cite{JaKa95} is called {\it multibondic}.

\subsection{How to get MUCA results?}

The aim is to sample a broad energy density, like the uniform
$$ P(E) = const. ~~{\rm for}~~ E_{\min} < E < E_{\max} $$
where $e_{\min}=E_{\min}/V$ and $e_{\max}=E_{\max}/V$ may be
kept constant. The uniform density is obtained by sampling, for 
$E_{\min} < E < E_{\max}$, with the weight factor
$$ w (E) = 1 / n(E) $$
where $n(E)$ is the spectral density. As $n(E)$ is a-priori
unknown, some preliminary or iterative estimate of $n(E)$ has 
to be part of the approach. For attempts to optimize the weight 
factors see \cite{HeSt95}.

\subsection{Weight factors and temperature}

Let us re-write the weight factor as 
$$ w(E) = e^{-S(E)} = e^{-\beta (E)\, E + \alpha (E)} $$
where $S(E)$ is the {\it microcanonical entropy}. Then,
the temperature follows from
$$ T(E) = 1/\beta (E) ~~{\rm and}~~ \beta (E) = 
{\partial S (E) \over \partial E} $$
Further, the function $\alpha (E)$ is determined up to 
an additive constant.
\bigskip

{\it Example with discrete energy:}

$$ \beta (E) = [S(E+\epsilon)-S(E)]/\epsilon ~~(\epsilon\
{\rm smallest\ energy\ step}) $$
The identity $S(E)=\beta(E)E-\alpha(E)$ implies
$$S(E)-S(E-\epsilon) = \beta(E)E - \beta(E-\epsilon)
(E-\epsilon) - \alpha (E) + \alpha (E-\epsilon) $$
Inserting $\epsilon \beta(E-\epsilon) = S(E) - S(E-\epsilon)$
yields the recursion \cite{BeNe91,BeNe92}
$$ \alpha(E-\epsilon) = \alpha (E) + [\beta(E-\epsilon) -
\beta(E)] E $$
where $\alpha (E_{\max}) =0$ is a convenient choice of the integration
constant.

\subsection{Recursive weight factor estimates}

For spin systems with first-order phase transitions the finite size
scaling (FSS) behavior is relatively well-known. Provided the steps
between system sizes are not too large, it is then possible to get
working estimates of the MUCA weights by means of FSS extrapolation
from the already simulated smaller systems \cite{BeHaNe93}. Another
method which works well for these systems is patching of overlapping,
constraint \cite{Bh87} MC simulations. While these approaches seem to
work well for static slowing down, they fail for the dynamic slowing down
of spin glass simulations. Then it is recommended to employ a more
sophisticated recursion \cite{Be96}, as outlined now.

It is advisable to start a recursion for MUCA weights in the 
{\it disordered region} (for which reliable canonical calculations can
be performed). For instance,
$$ \beta^0(E) \equiv \beta^0 (E_{\max}) ~~{\rm e.g.}~~
\beta^0(E_{\max}) = 0 $$
Then
\begin{equation}
\beta^{n+1}(E) = \epsilon^{-1}\, \ln [ \hat{H}^n_0 (E+\epsilon) /
\hat{H}^n_{\beta} (E) ]   \label{recursion}
\end{equation}
where $\hat{H}^n_x (E),\ (x=0,\beta)$ contains {\it combined}
information from the runs with $\beta^0(E), ..., \beta^n(E)$:
$$ \hat{H}^n_x (E) = \sum_{k=0}^n g^k(E)\, H^k_x (E) $$
Here $H^k_0(E)$ is the unnormalized histogram obtained from the
(short) simulation with $\beta^k(E)$. Further,
$$ H^k_{\beta} (E) = H^k_0 (E)\, e^{-\beta^k (E)\, \epsilon} $$
and the factors $g^k(E)$ weight the runs suitably. For instance,
$$ g^k(E) = const.\ \min [ H^k(E+\epsilon),\, H^k(E) ] $$
where the constant follows from the normalization $\sum_k g^k=1$.
For energy regions in which the statistics is (still) insufficient,
equation (\ref{recursion}) may be supplemented by
$$ \beta^{n+1} (E) = \beta^{n+1} (E+\epsilon)\, . $$
An interesting recent idea~\cite{SmBr96} is to utilize transition
frequencies instead of simple histograms. Finally, connections with
adaption and linear response theory have been explored~\cite{MuOy96}.

\subsection{Slowing down}

Our typical situation is
$$ E_{\max} - E_{\min} \sim V $$
The MUCA optimum for a flat energy distribution is given by a 
random walk in the energy. This implies a CPU time increase
$$ \sim V^2 $$
to keep the number of $E_{\max}\to E_{\min}\to E_{\max}$
transitions constant.  The recursion (\ref{recursion})
needs an additional $\sim V^{0.5}$ (optimum)
attempts to cover the entire range. It follows:
$$ {\rm slowing\ down}\ \sim\ V^{2.5}\ {\rm or\ worse.}$$
{\it Recursion alternative} (patching of overlapping
constraint \cite{Bh87} MC simulations):
$$ {\rm number\ of\ (fixed\ size)\ patches}\ \sim\ V\, .$$
When results can be obtained by keeping the number of updates per
spin (sweeps) in each patch constant, another CPU factor $\sim V$
follows. In this case we can get:
$$ {\rm optimal\ performance}\ \sim\ V^2\, .$$
This is still the optimal slowing down when the MUCA parameters can 
be obtained via FSS extrapolations.

\subsection{Related methods}

Combining MUCA with multigrid methods has been explored in 
Ref.\cite{JaSa94} and the cluster version~\cite{JaKa95} has already been 
mentioned. For molecular dynamics, Langevin and hybrid MC variants
see Ref.\cite{HaOkEi96}.

The {\it method of expanded ensembles} \cite{Ly92} proposes to
enlarge the configuration space by introducing new dynamical
variables. In {\it simulated tempering} \cite{MaPa92} it is the
temperature. A discrete set of weight factors is introduced
$$ w_k = e^{-\beta_k E + \alpha_k},\ k=1, ..., n,\
   \beta_1 < \beta_2 < ... < \beta_{n-1} < \beta_n $$
The transitions
$$ (\beta_k,\alpha_k) \to (\beta_{k-1},\alpha_{k-1}),\,
(\beta_{k+1},\alpha_{k+1}) $$
are now added to the usual $E\to E'$ transitions.
\bigskip

{\it Remarks:}

\begin{itemize}
\item The method works for dynamic but not for static 
supercritical slowing down, because each member of the discrete
set of weight factors samples still a Boltzmann distribution
(e.g. it is not suitable for calculating interfacial tensions).
\item In the context of massively parallel computer architectures
(and beyond) the variant of {\it parallel tempering} is 
particularly promising \cite{HuNe95}.
For additional information see the article by Marinari in these 
proceedings.
\end{itemize}

{\it Random Cost} \cite{Be93} sacrifices the exact relationship with
the canonical ensemble in favor of having a-priori defined transition 
probabilities. Assume, we can choose from a discrete set of updates, such
that an update implies one of the following energy changes:
$$ \triangle E^+_i,\, (i=1,...,n^+),\ \triangle E^0_j,\, (j=1,...,n^0)
 ~{\rm or}~ \triangle E^-_k,\ (k=1,...,n^-) $$
where $\triangle E^+_i>0$, $\triangle E^0_j=0$ and $\triangle E^-_k<0$.
It is then easy to define update probabilities $p^+_i$, $p^0_j$ and $p^-_k$,
$\sum_i p^+_i + \sum_j p^0_j + \sum_k p^-_k =1$, such that
\begin{equation}
\sum_i p^+_i \triangle E^+_i = - \sum_k p^-_k \triangle E^-_k \label{RC}
\end{equation}
holds (but at the extrema). The algorithm performs a random walk in 
the energy and, hence, samples a broad energy distribution. It is
conjectured to be of advantage in optimization problems, where one is 
mainly interested in minima and less in the canonical ensemble. 

\section{First-Order Phase Transitions}

MUCA simulations are best established for investigations of first-order 
phase transitions. The range of applications goes from MUCA studies of
mathematically ambitious topics to chemistry oriented ones. A MUCA 
investigation~\cite{BiNe93} of the Borgs-Koteck\'y \cite{BoKo91} FSS theory 
shows that very strong phase transitions or very large lattices are needed 
to observe the asymptotic behavior. To give two examples from the chemistry 
side, an investigation of the coexistence curve of the Lennard-Jones fluid 
was performed in Ref.\cite{Wi95} and the liquid-vapor asymmetry in pure
fluids was studied in Ref.\cite{WiMu95}. 

Most work has focused on calculations of interfacial tensions and an
over\-view is given in the forthcoming.

\subsection{$2d$ Potts Models}

A pioneering MUCA  study was performed for the $2d$ ten-state Potts
model~\cite{BeNe92}. Using the histogram method~\cite{Bi82}, the value
$2 f^s = 0.0978 (8)$ was found through FSS study of the equation
\begin{equation}
2 f^s_L = - {1\over L} \ln P^{\min}_L ~~{\rm where}~~ P_L^{\min} 
= \min \{ P_L(E)\, |\, E_L^{\max,1}<E<E_L^{\max,2} \}\, . \label{hist}
\end{equation}
Here $P_L(E)$ is the normalized energy density at the temperature
defined by $P_L^{\max,1} = P_L^{\max,2}$. When the first MUCA estimate
was published, numerical estimates of Potts model interfacial tensions
disagreed up to one order of magnitude. But, shortly after
the exact value was found to be $2f^s = 0.094701...$, 
see~\cite{BuWa93,BoJa92} and references therein. Certainly, this 
helped towards the break-through of MUCA methods. Once, the exact result
was known, the remaining, small discrepancy could be eliminated by
improving the finite volume estimators~\cite{BiNe94}. 

For these simulations the MUCA slowing down is around $\sim V^{2.3}$, 
{\it i.e.} reasonably close to the optimal performance.
For related investigations of the seven-state $2d$ Potts model, see
Ref.\cite{JaBe92,Ru93}.
On the technical side, arguments in favor of using equal weights
(instead of equal heights), when applying equation (\ref{hist}) to
asymmetric first-oder transitions, have been raised~\cite{BoKa92}.

\subsection{$2d$ and $3d$ Ising Model}

Many real physical systems fall into the universality class of the $3d$
Ising model. Despite its simplicity, it is therefore a very rewarding 
object to study.  Although many of its universal parameters have already
been determined with high precision, accurate results for some are still 
in the making. In particular, there has been recent interest in the
universal surface tension and the critical-isotherm amplitude 
ratios~\cite{ZiFi96}. 
To obtain them, one needs accurate interfacial tension results below the Curie
temperature. Here multimagnetical (MUMA) simulations~\cite{BeHaNe93}
have become the enabling technique for Binder's~\cite{Bi82} histogram
method, which was originally proposed in this context.

For the $2d$ Ising model Onsager's exact result could be reproduced with
good accuracy. However, for the $3d$ model the temperature dependence
of MUMA interfacial tensions has come out fairly erratic. Therefore, the
results of Ref.\cite{HaPi94} should be regarded as the up-to-date best
estimates. Meanwhile, considerable technical improvements of MUMA
calculations are feasible (see also the next subsection) and it
seems worthwhile to start off a new, large scale, MUMA $3d$ Ising model
simulation.

\subsection{$SU(3)$ Gauge Theory}

One is interested in the interfacial tension for the
confinement/deconfinement phase transition. The use of MUCA techniques
has been explored by Grossmann, Laursen et al. \cite{GrLa92,GrLa93}.
In particular, they noticed that it is suitable to use an asymmetric
lattice, $V=L_z L^2 L_t$ with $L_z\ge 3L$. This forces the interfaces
into the $L^2$ plane and ensures a flat region for the minimum of 
equation  (\ref{hist}), thus greatly facilitating the extraction of 
finite-lattice values for the interfacial tension and, consequently,
the FSS analysis. 

For $SU(3)$ gauge theory the interfacial tension is
usually denoted by the symbol $\sigma$ and estimates are conveniently
given as multiples of $T_c^3$, where $T_c$ is the deconfinement 
temperature. Using the conventions of \cite{IwKa94}, the estimates
of~\cite{GrLa93} seem to be
$$ \sigma = 0.052 (4)\, T_c^3,\ (L_t=2) ~~{\rm and}~~ 
   \sigma= 0.020 (2)\, T_c^3,\ (L_t=4) $$
This may be compared with the later estimate by Iwasaki et al. \cite{IwKa94}
$$ \sigma = 0.02925 (22)\, T_c^3,\ (L_t=4) ~~{\rm and}~~ 
   \sigma= 0.0218 (33)\, T_c^3,\ (L_t=6) $$
The discrepancy (only $L_t=4$ can be compared) is presumably due to
too small lattice sizes in the MUCA study. Physically, one is interested
in the $L_t\to\infty$ limit. Recently, it has been suggested that the
the strong $L_t$ dependence can be eliminated by using tadpole improved 
actions. Beinlich et al. \cite{BeKa96} report
$$ \sigma = 0.0155 (16)\, T_c^3,\ (L_t=2 ~~{\rm\bf and}~~ L_t=3)\, .$$

\subsection{Electroweak Phase Transition}

Baryon violating processes are unsuppressed for $T>T_c$, where $T_c$
is the electroweak critical temperature. It has been conjectured, that
this may allow to explain the baryon asymmetry in nature. Models tie
the nucleation rate to the interface tension of the transition. Using
an effective scalar field theory~\cite{KaNePa95} or the full
theory~\cite{CsFo95}, MUCA and related techniques turn out to be useful 
for simulations at a Higgs mass
$$ m_H \approx (35-37)\, GeV\, , $$
where one deals with a relatively strong first-order transitions, as
needed to explain the baryon asymmetry. Unfortunately, it turn out that
the transition weakens for higher Higgs masses, see Ref.\cite{Mo95}
for a concise review.

\section{Systems with Conflicting Constraints}

In these systems one encounters large free energy barriers due to
disorder and frustrations. MUCA simulations try to overcome the
barriers through excursions into the disordered phase. Examples are
spin glasses, proteins (see Hansmann in these proceedings),
hard optimization problems and others.

\subsection{Spin glasses}

MUCA studies have so far focused on the simplest,
not-trivial prototypes, the $2d$ and $3d$ Edwards-Anderson Ising (EAI) 
spin glass~\cite{BeCe92,BeHaCe94}. Significant progress is achieved with
respect to groundstate energy {\it and} entropy calculations. However,
the slowing down with volumes size is very bad, around $V^4$ or,
possibly, exponential. Certain advantages of simulated tempering are
claimed in Ref.\cite{KeRe94}. An uncontroversial one is that the latter 
approach can easily be vectorized, whereas other issues have 
straightforward (but yet untested) translations into the MUCA approach. 
In any case, presently we have no indication that simulated 
tempering yields a significantly improved slowing down.

\subsection{Optimization problems}

They occur in engineering, network and chip design, traffic control, and
many other situations. General purpose algorithms for their solution are 
Simulated Annealing and Genetic Algorithms. To those, we may now like to 
add Multicanonical Annealing~\cite{LeCh94} and Random Cost~\cite{Be93}.

Multicanonical Annealing is a combination of MUCA sampling with variable
upper bounds and frequent adaption of parameters. A promising 
study~\cite{LeCh94} has been performed for the traveling salesman
problem. Up to $N=10,000$ cities, randomly distributed in the unit square,
were considered and scaling of the path length as function of $N$ was
investigated. The reported algorithmic performance is so good, that an 
independent confirmation would be highly desireable.

Two non-trivial applications of the random cost algorithm emerged
recently: (i)~It was used for training a feed-forward
multilayer perceptron, relevant for analyzing high energy physics 
experimental data. (ii)~Topology optimization as needed for the engineering
of trusses was studied~\cite{Ba96}.

\section{Conclusions}

Sampling of broad energy distributions allows to overcome supercritical
slowing down. This is well established for first-order phase transitions.
Systems with conflicting constraints remain, despite some progress,
notoriously difficult and for them most hope lies on achieving further
algorithmic improvements. Finally,
MUCA methods may also be of interest when dealing with second order phase 
transitions, but so far little experience exists in this direction.

\section*{Acknowledgments}
This work was partially supported by the Department of Energy under
contract DE-FG05-87ER40319. I would like to thank ZIF and Frithjof
Karsch for their hospitality and J\"urgen Riedler for comments on the
manuscript.

\end{document}